\documentclass[12pt]{article}
\usepackage{epsfig,cite}
\usepackage {amsmath}
\usepackage{amssymb}
\usepackage{color}
\usepackage{graphicx}
\usepackage{verbatim,mathrsfs,subfig}
\include{epsf}
\newcount\timecount
\newcount\hours \newcount\minutes  \newcount\temp \newcount\pmhours
\hours = \time
\divide\hours by 60
\temp = \hours

\multiply\temp by 60
\minutes = \time
\advance\minutes by -\temp
\def\hour{\the\hours}
\def\minute{\ifnum\minutes<10 0\the\minutes
            \else\the\minutes\fi}
\def\clock{
\ifnum\hours=0 12:\minute\ AM
\else\ifnum\hours<12 \hour:\minute\ AM
      \else\ifnum\hours=12 12:\minute\ PM
            \else\ifnum\hours>12
                 \pmhours=\hours
                 \advance\pmhours by -12
                 \the\pmhours:\minute\ PM
                 \fi
            \fi
      \fi
\fi
}

\def\monthname{\relax\ifcase\month 0/\or January\or February\or
   March\or April\or May\or June\or July\or August\or September\or
   October\or November\or December\else\number\month/\fi}

\def\bold#1{\setbox0=\hbox{$#1$}%
     \kern-.025em\copy0\kern-\wd0
     \kern.05em\copy0\kern-\wd0
     \kern-.025em\raise.0433em\box0 }

\def\beq{\begin{equation}}
\def\eeq{\end{equation}}

\def\ga{\mathrel{\raise.3ex\hbox{$>$\kern-.75em\lower1ex\hbox{$\sim$}}}}
\def\la{\mathrel{\raise.3ex\hbox{$<$\kern-.75em\lower1ex\hbox{$\sim$}}}}
\def\gev{{\rm \, Ge\kern-0.125em V}}
\def\tev{{\rm \, Te\kern-0.125em V}}
\def\gyr{{\rm \, G\kern-0.125em yr}}

\def\gappeq{\mathrel{\rlap {\raise.5ex\hbox{$>$}}
{\lower.5ex\hbox{$\sim$}}}}
\def\lappeq{\mathrel{\rlap{\raise.5ex\hbox{$<$}}
{\lower.5ex\hbox{$\sim$}}}}
\def\Toprel#1\over#2{\mathrel{\mathop{#2}\limits^{#1}}}

\def\m12{m_{1\!/2}}

\def\bea{\begin{eqnarray}}
\def\eea{\end{eqnarray}}

\def\beq{\begin{equation}}
\def\eeq{\end{equation}}

\begin{document}

\begin{titlepage}
\pagestyle{empty}
\baselineskip=21pt
\rightline{UMN--TH--3328/14, FTPI--MINN--14/7, IPMU14-0040}
\vspace{0.2cm}
\begin{center}
{\large {\bf Peccei-Quinn Symmetric Pure Gravity Mediation }}
\end{center}
\vspace{0.5cm}
\begin{center}
{\bf Jason L. Evans}$^{1}$,
{\bf Masahiro Ibe}$^{2,3}$ {\bf Keith~A.~Olive}$^{1}$
and {\bf Tsutomu T. Yanagida}$^{3}$\\
\vskip 0.2in
{\small {\it
$^1${William I. Fine Theoretical Physics Institute, School of Physics and Astronomy},\\
{University of Minnesota, Minneapolis, MN 55455,\,USA}\\
$^2${ ICRR, University of Tokyo, Kashiwa 277-8582, Japan}\\
$^3${Kavli IPMU (WPI), TODIAS, University of Tokyo, Kashiwa 277-8583, Japan}\\
}}
\vspace{1cm}
{\bf Abstract}
\end{center}
\baselineskip=18pt \noindent
{\small }
Successful models of Pure Gravity Mediation (PGM) with radiative
electroweak symmetry breaking can be expressed with as few as two free parameters
which can be taken as the gravitino mass and $\tan \beta$.  These models
easily support a 125-126 GeV Higgs mass at the expense of a scalar spectrum in the multi-TeV
range and a much lighter wino as the lightest supersymmetric particle.
In these models, it is also quite generic that the Higgs mixing mass parameter, $\mu$,
which is determined by the minimization of the Higgs potential is also in the multi-TeV range.
For $\mu > 0$, the thermal relic density of winos is too small to account for the dark matter.
The same is true for $\mu < 0$ unless the gravitino mass is of order 500 TeV.
Here, we consider the origin of a multi-TeV $\mu$ parameter arising from the breakdown
of a Peccei-Quinn (PQ) symmetry. A coupling of the PQ-symmetry breaking field, $P$,
to the MSSM Higgs doublets, naturally leads to a value of $\mu \sim \langle P \rangle^2 /M_P \sim
\mathcal{O}(100)$ TeV and of order that is required in PGM models.  In this case,
axions make up the dark matter or some fraction of the dark matter with
the remainder made up from thermal or non-thermal winos. We also provide solutions
to the problem of isocurvature fluctuations with axion dark matter in this context.

\vfill
\leftline{February 2014}
\end{titlepage}

\section{Introduction}
The mass of the Higgs boson\cite{lhch} at around $126$\,GeV is near the upper limit of the predictions
in commonly studied models of the supersymmetric standard model such as the
constrained minimal supersymmetric standard model (CMSSM)
\cite{cmssm,cmssmh}~\footnote{The difficulty in obtaining a 126 GeV Higgs in the CMMSSM
has been relaxed recently with the inclusion of higher order corrections to the Higgs mass
calculation \cite{relax,ehow+}}.
This rather large Higgs boson mass and the null results of the sparticle searches  at the LHC \cite{lhc,ATLASsusy}
seem to hint to rather heavy sparticles \cite{post-LHC,hope}, and possibly heavy top squarks with masses in the range of  tens to hundreds of TeV.

Among the models with heavy sparticles, models with
a mild hierarchy between sfermion and
gaugino masses such as  pure gravity mediation (PGM) \cite{pgm,pgm2,pgm3,eioy,eioy2,eioy3}
and models with strongly stabilized moduli\cite{klor,Dudas:2006gr,dlmmo} are very successful not only
phenomenologically but also cosmologically and have spectra which are characteristic of split
supersymmetry \cite{split} with anomaly mediation \cite{anom}.
In these models, the sfermions obtain tree-level masses of the order of the gravitino mass, $m_{3/2}$,
while the gaugino masses are dominated by one-loop masses from anomaly mediation\cite{anom}.

In the original version of PGM\cite{pgm},
it was assumed that the supersymmetric Higgs mixing term,
the $\mu$-term, is generated via the tree-level
couplings to an $R$-symmetry breaking sector which generates the non-vanishing vacuum
expectation value of the superpotential\,\cite{gm}.
In this case, the $\mu$ term and the supersymmetry breaking bilinear $B$ term are two
independent parameters and are both of order the gravitino mass, $m_{3/2}$.
In previous papers, we showed that models based on PGM with \cite{eioy} and without \cite{eioy2}
scalar mass universality, could explain virtually all experimental constraints with successful radiative electroweak symmetry breaking (EWSB). In this case, one can impose the supergravity boundary
condition $B_0 = A_0 - m_{3/2}$ which is essentially $-m_{3/2}$ since $A_0$ is also determined
by anomalies and hence small compared with $m_{3/2}$.
The universal scalar mass, $m_0$ is fixed by $m_{3/2}$. The $\mu$ term is determined by
the minimization of the Higgs potential. The ratio of the Higgs vacuum expectation
values (vevs) can be treated as a free parameter so long as a Giudice-Masiero coupling, $c_H$,
is included in the K\"ahler potential, $K$. The value of $c_H \lesssim 1$ is then also fixed by the minimization
of the Higgs potential.

In this paper, we discuss another version of PGM, in which the $\mu$-term originates from the breaking
of a Peccei-Quinn (PQ) symmetry \cite{PQ} via a dimension five operator\cite{KN}.
The size of the $\mu$-term is determined by the PQ-breaking scale, $f_{PQ}$,
and hence, it can be related to the axion dark matter density. In practice, the $\mu$ term
and $f_{PQ}$ are then determined by the EWSB boundary conditions at the weak or susy
scale. As in \cite{eioy}, the $B$-term is fixed to $-m_{3/2}$ at the input UV scale.
However, because the Higgs fields carry PQ charge in this model, a Giudice-Masiero coupling
is not allowed in $K$ and therefore some departure of scalar mass universality is required \cite{eioy2}.
With $c_H = 0$, $\tan \beta$ must also
be determined by the EWSB boundary conditions at the weak or susy scale.
We will explore a three parameter version of the PGM where the three free parameters are
chosen to be the gravitino mass, $m_{3/2}$, and the two soft Higgs masses, $m_1$ and $m_2$.

We show further that successful phenomenological models can be constructed
if only the soft mass of the up-type Higgs, $m_2$,  is non-universal.
Thus the family of PQ symmetric PGM models can be expressed in terms of only
two parameters, $m_{3/2}$ and $m_2$. In fact, viable solutions are possible
with $m_1 = m_{3/2}$ and $m_2 = 0$. Thus, if the up-type Higgs multiplet is
associated with a pseudo Nambu-Goldstone boson \cite{Kugo:1983ai} or if the Higgs soft mass is
protected by a (partial) no-scale structure \cite{noscale} of the K\"ahler potential as discussed in \cite{eioy2}, we are reduced to a theory with one single free parameter!
Finally we will show that an alternative model for breaking the PQ-symmetry based on \cite{Murayama:1992dj} may allow full universality to be restored.

As in previous studies of PGM, we also find that the lightest supersymmetric particle (LSP) is the neutral wino. 
In the $R$-parity conserving case, the model allows several dark matter scenarios, including one with pure wino dark matter
with possible non-thermal sources such as the late time decay of the gravitino.
As we will see, even if the wino contribution to the relic density is negligible
(as is the case for the thermal contribution when $\mu > 0$), it is possible
that axions make up the entire dark matter, or of course it is also possible that
the dark matter is an axion-wino mixture. In the scenario where a significant portion of the dark matter is the axion,  the wino can be much lighter than the $3$ TeV as required by thermal dark matter constraints. In fact, it could have a mass below $1$ TeV and  fall within the reach of the LHC at 14 TeV \cite{pgm3}. While it is well known that models with axion
dark matter generally overproduce isocurvature fluctuations \cite{iso}, it is also known
that the problem may be resolved if the axion decay constant takes on
values during inflation which are large compared to the nominal low energy value \cite{al91}.
We show several ways, this can be implemented in the models presented.

The organization of the paper is as follows.
In section \ref{sec:mu}, we summarize PGM  with universal and non-Universal Higgs masses.
We then generalize the model to include a PQ-symmetry in which the $\mu$-term originates from
PQ-symmetry breaking.
In section \ref{sec:PQ}, we briefly review the properties of axion dark matter.
There, we also propose several solutions to the problem of
isocurvature fluctuations including a novel model of dynamical PQ-breaking.
In section \ref{sec:NU}, we demonstrate that this version of PGM achieves successful
EWSB. Here, we will consider the case with two non-universal Higgs soft masses,
that is the three parameter version of the model.
We will see however, that the down-type Higgs soft mass may remain universal,
and successful models are still obtained. We will further demonstrate that even in the
special case where $m_2 = 0$, viable models are possible so long as $300$ TeV $\lesssim m_{3/2} \lesssim 850$ TeV.
Before concluding, we will briefly describe in section \ref{sec:pquniv} an alternative model for breaking
the PQ-symmetry in which full universality may be restored.
The final section is devoted to discussions.

\section{$\mu$-term from $PQ$-symmetry breaking}\label{sec:mu}
\subsubsection*{Pure Gravity Mediation}
In pure gravity mediation models, it is assumed that all fields in the
supersymmetry (SUSY) breaking sector are charged under some symmetry (e.g. an R-symmetry).
Under this assumption, the gaugino masses and the $A$-terms of the chiral multiplets
are suppressed at the tree-level in supergravity and they are dominated by the anomaly mediated contributions\,\cite{anom}.
The soft squared masses of the scalar bosons are, on the other hand, generated at the tree-level and
are expected to be of order $m_{3/2}^2$ with a generic K\"ahler potential.
If we optionally assume a flat K\"ahler manifold for all the MSSM fields, all of the MSSM scalar fields
obtain the universal soft mass squared equal to $m_{3/2}^2$.
In our analysis, we take the model with universal scalar masses as our starting point.
We will assume universality at the grand unified (GUT) scale and run all quantities down
to the weak scale through 2-loop RGEs and minimize the Higgs potential. As in the CMSSM,
minimization allows us to solve for $\mu$ and the bilinear supersymmetry breaking $B$-term.
As we will see in our later analysis, we will be required to consider a slightly relaxed assumption
where the K\"ahler manifold is flat
for all the MSSM field except for the Higgs doublets, which leads to non-universal Higgs soft masses (NUHM) \cite{nonu,nuhm2,nuhm1}.

In the original PGM model, we assumed that the holomorphic bi-linear of the two-Higgs doublets, $H_uH_d$, has a vanishing $R$-charge
and appears in the K\"ahler potential and the superpotential,
\begin{eqnarray}
\label{eq:K1}
  K|_{H_uH_d} &=& c_H H_u H_d + h.c.\ , \\
  \label{eq:W1}
  W|_{H_uH_d} &=& c_H'm_{3/2} H_u H_d \ .
\end{eqnarray}
where $c_H$ and $c_H'$ are $O(1)$ coefficients\cite{gm}.
From these terms, the $\mu$ and the $B$-parameters are given by,
\begin{eqnarray}
\label{eq:mu1} \mu &=& c_H m_{3/2} + c_H' m_{3/2}\ , \\
 \label{eq:B1}B\mu &=& 2 c_H m_{3/2}^2 - c_H' m_{3/2}^2  \ .
\end{eqnarray}
In minimal supergravity, we would take $c_H = 0$, so that
$\mu = c_H' m_{3/2}$ and $B  = -m_0$, neglecting the small contribution
from anomaly mediated $A$-terms. Minimization of the Higgs potential in this case
allows one to solve for $\mu$ ($c_H'$) and $\tan \beta$.
 Inclusion of $c_H$, allows one to keep
$\tan \beta$ as a free parameter solving instead for $\mu$ ($c_H'$) and $c_H$ \cite{dmmo}.
Thus, there are two independent parameters which can be chosen as $m_{3/2}$ and $\tan \beta$.%
\footnote{
The $B\mu$ term may also obtain comparable contributions from higher dimensional operators
in the K\"ahler potential,
\begin{eqnarray}
{\mit \Delta}K|_{H_uH_d} = \frac{Z^\dagger Z}{M_P^2} H_u H_d + h.c.
\end{eqnarray}
where $Z$ denotes the supersymmetry breaking field.
}

\subsubsection*{PQ-symmetric PGM}
Now, let us move on to the PQ-symmetric PGM model where the $\mu$-term is generated by the breaking of the PQ-symmetry.
In this case, instead of the above potentials in Eqs.\,(\ref{eq:K1}) and (\ref{eq:W1}), the $\mu$-term is generated via,
\begin{eqnarray}
W|_{H_uH_d} =  k \frac{P^2}{M_P} H_u H_d\ ,
\label{wpq}
\end{eqnarray}
where $k$ denotes a dimensionless constant and $P$ is a PQ-symmetry breaking field
with PQ-charge $+1$. $M_P =
2.4 \times 10^{18}$ GeV refers to the reduced Planck scale.
We assumed that the holomorphic bi-linear $H_uH_d$ has a PQ charge of $-2$ and
the  charges of  other MSSM matter fields are  appropriately assigned.  As a result,
the K\"ahler term in Eq. (\ref{eq:K1}) is not allowed.

From this potential, we obtain the $\mu$ and the $B$-parameters,
\begin{eqnarray}
\label{eq:mu2}
 \mu &=&k \frac{\langle{P}\rangle^2}{M_P} \ ,  \\
 \label{eq:B2}
 B\mu &=& - m_{3/2}\mu\ .
 \end{eqnarray}
Therefore, the $\mu$-parameter is determined by the PQ-breaking scale and, for example, $\mu = {\cal}O(100)$\,TeV
is realized for $\langle P \rangle = {\cal O}(10^{12})$\,GeV.
It should be noted that $\mu$ cannot be much larger than the gravitino mass for successful EWSB\footnote{EWSB requires either $m_{\tilde t}^2\sim \mu^2$ or $B\mu\sim \mu^2$. In either case, we get $\mu\sim m_{3/2}$.}.
The $B$-parameter is, on the other hand, fixed to $-m_{3/2}$ (at the GUT scale),
which should be contrasted
to the above original PGM where the $B$-parameter (through $c_H$) is an independent parameter
which for convenience can be exchanged with $\tan \beta$ as described above.
As we will discuss later, this restricted parameter set conflicts with full scalar mass universality.

In summary, the PQ-symmetric PGM model is more restricted  than the original PGM model and has
effectively only one parameter
\begin{eqnarray}
m_{3/2}
\end{eqnarray}
 with full scalar mass universality. Since $B$ is fixed, the EWSB conditions
 amount to solving for $\tan \beta$ and the constant $k$ in Eq. (\ref{wpq}), which has no solution.
Here, we have traded the size of $\mu$-term with the $Z$-boson mass, fixed to $m_Z \simeq 91.2$\,GeV.
Even with the NUHM2 \cite{nuhm2}, the model has only three parameters,
\begin{eqnarray}
m_{3/2}, \quad m_1,  \quad m_2 ,
\end{eqnarray}
where $m_{1,2}$ denote the soft masses of the two Higgs doublets,
$H_d, H_u$. However, as we
shall see, viable solutions are possible with non-universality extended only to the up-type
Higgs, ie. we will be able to keep $m_1 = m_{3/2}$.  Furthermore, we will also see that the
special case of $m_2 = 0$ also yields acceptable solutions, so if $H_2$ originates as a
pseudo Nambu-Goldstone boson, or $H_2$ is part of a no-scale structure in $K$,
we are again reduced to a one-parameter theory.

\section{Supersymmetric axion model}
\label{sec:PQ}
\subsubsection*{Brief review of the PQ-breaking model}
Before going on to study the parameter space for the PQ-symmetric PGM model,
let us discuss the supersymmetric axion model\,\cite{axion} in a little more detail.
As a simple example, we may take the following model of spontaneous PQ-symmetry breaking,
\begin{eqnarray}
W_{\rm PQ} = \lambda X(PQ - v_{PQ}^2)\ ,
\label{wpq2}
\end{eqnarray}
where $\lambda$ is a dimensionless coupling, and $v_{PQ}$ a dimensionful parameter.
The superfields $X$, $P$, and $Q$ have PQ-charges, $0$, $+1$ and $-1$, respectively,
with vacuum values,
\begin{eqnarray}
\langle P \rangle = \langle Q \rangle \simeq v_{PQ}\ , \quad
\langle X \rangle \simeq m_{3/2} / \lambda\ ,
\end{eqnarray}
so that the PQ-symmetry is spontaneously broken,%
\footnote{Here, we are assuming that the soft squared masses of $P$ and $Q$ are equal to each other.
When the soft squared masses of $P$ and $Q$ are quite different from each other,
the $B$-parameter in Eq.(\ref{eq:B2}) obtains a sizable correction\,\cite{NT,HIY}.
}
and we are left with an axion, saxion and axino at an energy scale much lower than $v_{PQ}$.

Due to SUSY breaking effects, the saxion and the axino obtain masses of order of the gravitino mass,
and hence, they decay quickly to a pair of the gluinos and a gluino/gluon pair, respectively,
and cause no cosmological problems.
The axion, on the other hand, remains very light and has a very long lifetime, and
can be a good dark matter candidate with a relic density\,\cite{Turner:1985si},
\begin{eqnarray}
\Omega_ah^2 = 0.18\left( \frac{F_{PQ}}{10^{12}\,\rm GeV}\right)^{1.19}\left(\frac{\Lambda}{400\,\rm MeV}\right)
\label{relic}
\end{eqnarray}
where $\Lambda$ denotes the QCD scale and the decay constant $F_{PQ}$ is defined by $v_{PQ}$ and determined
by a domain wall number  $N_{DW} = 6$ in this case.
We assumed the initial axion amplitude $a \simeq F_{PQ}$.
Thus, the axion can be the dominant dark matter component for $F_{PQ}\simeq 7 \times 10^{11}$\,GeV,
assuming that the mis-alignment angle is of order $\pi$.

It should be noted that  axion dark matter models are suspect to
problems with domain walls and isocurvature fluctuations.
The former problem is, however, solved relatively easily
if  the PQ-symmetry is broken before the primordial inflation starts
and is never restored after the end of inflation.
In this case, the domain walls are not formed after inflation, and hence,
there is no domain wall problem.

The latter problem, the overproduction of isocurvature fluctuations \cite{iso}, on the other hand,
puts a severe constraint on the Hubble scale during inflation.
In fact, in order to suppress the isocurvature mode in the axion dark matter enough to be consistent
with the constraint set by CMB observations \cite{planck},
the Hubble scale during inflation is required to be rather small (i.e. $H \lesssim 2 \times 10^7$GeV),
which is much smaller than the one in more conventional models of inflation where $H\sim 10^{13}$\,GeV.

It is, however, possible to relax the constraint on $H$, if the axion decay constant, $F_{PQ}$,
were larger than its low energy value during inflation \cite{al91}.
Roughly, we would require $H/F_{PQ} \lesssim 3 \times 10^{-5}$ to resolve the problem,
and hence, $F_{PQ} \gtrsim 3 \times 10^{17}$\,GeV during inflation for $H\sim 10^{13}$\,GeV.
Such a large PQ-breaking scale during inflation can be achieved, for example, when
$P$ in the above model picks up a negative mass squared contribution along the $PQ$ flat direction,
in a similar way that Affleck-Dine fields \cite{AD} pick up large vevs to generate
a baryon asymmetry (for a concrete model in this context see \cite{go}).

The PQ-breaking scale can also be large during inflation when the
radial component of the PQ-breaking field, $\sigma \equiv P = Q$
flows into the so-called attractor solution during inflation \cite{al91}.
In this case, the PQ-breaking field scales as $\lambda^{-1/2}$ where $\lambda$ is
the coupling in Eq. (\ref{wpq2})\cite{hiky}.
This may require couplings as small as $10^{-8}$ to $10^{-12}$
depending on the particular model of inflation. Because the amplitude of the $P$ and $Q$ fields are large during inflation, it is possible that subsequent large oscillations could pass through $P=Q=0$ which would amount to the restoration of the PQ-symmetry and could lead to domain wall formation. To determine if domain walls form, a detailed analysis of the relaxation of the $P$ and $Q$ fields is needed. It is important to note that the amount of relaxation depends on the model of inflation. For inflation models quadratic in the inflaton, the PQ fields relax too slowly and domain walls are formed, see \cite{kyy}.

It is also possible to relax this problem by considering a Giudice-Masiero-like term
involving the PQ fields in the K\"ahler potential,
\begin{eqnarray}
K\supset -c_{PQ}\left(PQ+P^\dagger Q^\dagger\right).\label{cpq}
\end{eqnarray}
Since the product of $P$ and $Q$ have $R$ and $PQ$ charge zero, this additional term cannot be forbidden. During inflation, the K\"ahler potential now gives a correction to the scalar potential
\begin{eqnarray}
\Delta V = -c_IH^2c_{PQ}\left(PQ+P^\dagger Q^\dagger\right)+c_IH^2\left(|P|^2+|Q|^2\right),
\end{eqnarray}
where $c_I \approx 3$ and $H$ is Hubble's constant.  Adding this to the scalar potential for the $PQ$ fields from the superpotential, we find
\begin{eqnarray}
V=|\lambda|^2\left|PQ-\frac{c_{PQ}c_I}{|\lambda|^2}H^2-\Lambda^2\right|^2 +c_IH^2\left(|P|^2+|Q|^2\right)+V_0,
\end{eqnarray}
where $V_0$ keeps track of the constant pieces in the potential. To get PQ breaking during inflation, we need $c_{PQ}>1$.  During inflation, the new minimum is around
\begin{eqnarray}
P=Q \sim \left( \frac{c_{PQ}c_I}{|\lambda|^2} \right)^{1/2} H.
\label{pqinf}
\end{eqnarray}
To sufficiently suppress the isocurvature perturbations, we need a somewhat small coupling,
\begin{eqnarray}
 \frac{|\lambda|}{\sqrt{c_{PQ}c_I}}\lesssim 3 \times 10^{-5},
 \end{eqnarray}
which is effectively a constraint of $\lambda \lesssim 10^{-4}$.

In this scenario, after inflation the $P$ and $Q$ fields will relax back to the true minimum set by $\Lambda$.  The masses of the $P$ and $Q$ fields during inflation are of order $\sqrt{(c_{PQ} - 1) c_I} H$. Therefore,  during their relaxation back to the true minimum, $P$ and $Q$ will track the minimum
(so long as $c_{PQ}$ is not very close to 1) and no domain walls will be formed, that is, the PQ-symmetry remains broken.

\subsubsection*{Dynamical PQ-symmetry breaking}
We may also consider an alternative mechanism for realizing a large PQ-breaking scale during inflation. If
PQ-breaking scale is generated dynamically it is possible to get this dynamical scale large during inflation.
To illustrate this mechanism, let us consider a supersymmetric $SU(2)$ gauge theory
with four fundamental chiral fields, $P_{1,2}$ and $Q_{1,2}$, and four singlet fields $S_{ij}\, (i,j=1\cdots2)$.
The superfields $P$, $Q$ and $S$ have PQ-charges $+1$, $-1$ and $0$, respectively,
and they are coupled through the superpotential,
\begin{eqnarray}
 W =
 \sum_{i,j=1,2}\lambda  S_{ij} P_{i} Q_{j}\ ,
\end{eqnarray}
where $\lambda$ is again a dimensionless coupling.
Below the dynamical scale of $SU(2)$, $\Lambda_{PQ}$, the model can be described by the six composite mesons,
\begin{eqnarray}
M_{PP} \simeq P_1P_2/\Lambda_{PQ}, \quad
M_{QQ} \simeq Q_1Q_2/\Lambda_{PQ}, \quad
{M_{PQ}}_{ij} \simeq P_iQ_j/\Lambda_{PQ},
\end{eqnarray}
whose effective superpotential terms are roughly given by,
\begin{eqnarray}
\label{eq:Weff}
 W_{\rm eff} \simeq \sum_{ij} \lambda \Lambda_{PQ} S_{ij} {M_{PQ}}_{ij} + X({\rm Pf}(M) - \Lambda_{PQ}^2 ) \ .
\end{eqnarray}
Here, $X$ denotes the Lagrange multiplier to impose the quantum deformed moduli constraint
on the mesons\,\cite{Seiberg:1994bz},
\begin{eqnarray*}
{\rm Pf}(M) = M_{PP} M_{QQ}  + \frac{1}{2} \sum_{}\epsilon^{ij}\epsilon^{kl} {M_{PQ}}_{ik}{M_{PQ}}_{jl} = \Lambda_{PQ}^2\ .
\end{eqnarray*}
By noting that the $M_{PQ}$ obtain masses of ${\cal O}(\lambda \Lambda_{PQ})$ from
the first term in Eq.\,(\ref{eq:Weff}), we find that the PQ-symmetry is spontaneously broken by,
\begin{eqnarray}
 \langle M_{PP}\rangle =
  \langle M_{QQ}\rangle \simeq \Lambda_{PQ}\ ,  \quad
  \langle X\rangle \simeq m_{3/2}\ , \quad
   \langle M_{PQ}\rangle = 0 \ ,
\end{eqnarray}
Therefore, this model is a dynamical realization of the previous model defined in Eq.\,(\ref{wpq2}).%
\footnote{This model is close to the model of dynamical PQ-breaking proposed in \cite{why}, where
the PQ-symmetry and supersymmetry are broken simultaneously.}

Now let us assume that the gauge coupling constant of $SU(2)$ is given by a vev of a singlet field $\phi$
through the gauge kinetic function, $f = (1/g_0^2 + \phi^2)$,
assuming $g_0 = 4 \pi$ and $\langle \phi \rangle=O(1) $ at the Planck scale.
Then the effective gauge coupling at the Planck scale is perturbative, $1/g^2=(1/g_0^2 + \langle\phi \rangle^2)$,
with $\langle\phi\rangle$ chosen so that the condensation scale is
$\Lambda_{PQ} \simeq 10^{12}$\,GeV.
However, if $\phi$ gets a large positive mass during inflation, the vev of $\phi$ will be suppressed, $\langle \phi\rangle \ll 1$.  Since the vev independent part of the gauge coupling is already strong at the Planck scale, we get $\Lambda_{PQ}=M_P$ during inflation.
In this way, we can easily realize a large PQ-breaking scale during inflation, and hence, the isocurvature fluctuations in axions can be suppressed. $\phi$ can be given a mass by adding the following interactions,
\begin{eqnarray}
W=\lambda^\prime Y\left(\phi^2-M_P^2\right).
\end{eqnarray}
At the minimum of the potential, $\phi$ has a Planck scale vev which generates a mass for $\phi$ of order $\lambda^\prime M_P$ avoiding a moduli problem~\footnote{ If $\phi$ gets a positive Hubble induced mass$^2$ $>(\lambda^\prime M_P)^2$, $\phi$ will be stabilized at its origin $\phi =0$.}.

\section{Results}
\label{sec:NU}

We are now in a position to explore the necessary parameter space
for the PQ-symmetric PGM model.  We begin with the three-parameter version of the model
defined by the gravitino mass, and the two soft Higgs masses. All other scalars
are assumed to be universal at the GUT scale (the renormalization scale where the
two electroweak couplings are equal). All masses and couplings are run down to the
weak scale, where the Higgs potential is minimized, thus determining $\mu$ (or $k$ in this
context) and $\tan \beta$. Gaugino masses and $A$-terms assume their
anomaly mediated values, and $B_0 = -m_{3/2}$. As we have
seen previously \cite{eioy2}, PGM solutions with $c_H = 0$ are possible
so long as we allow the Higgs soft masses to depart from universality.

\begin{figure}[h!]
\begin{minipage}{8in}
\epsfig{file=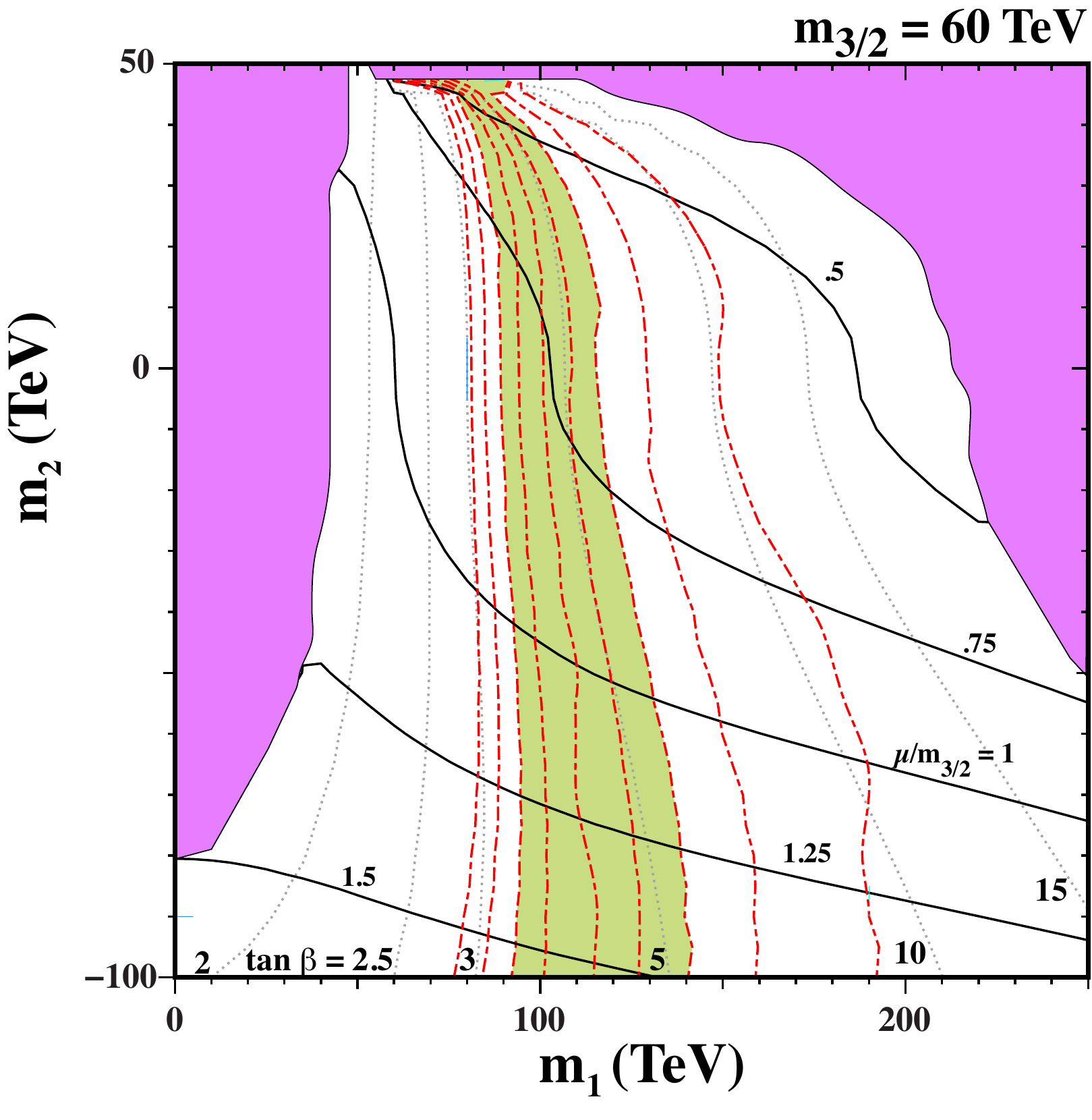,height=3.2in}
\epsfig{file=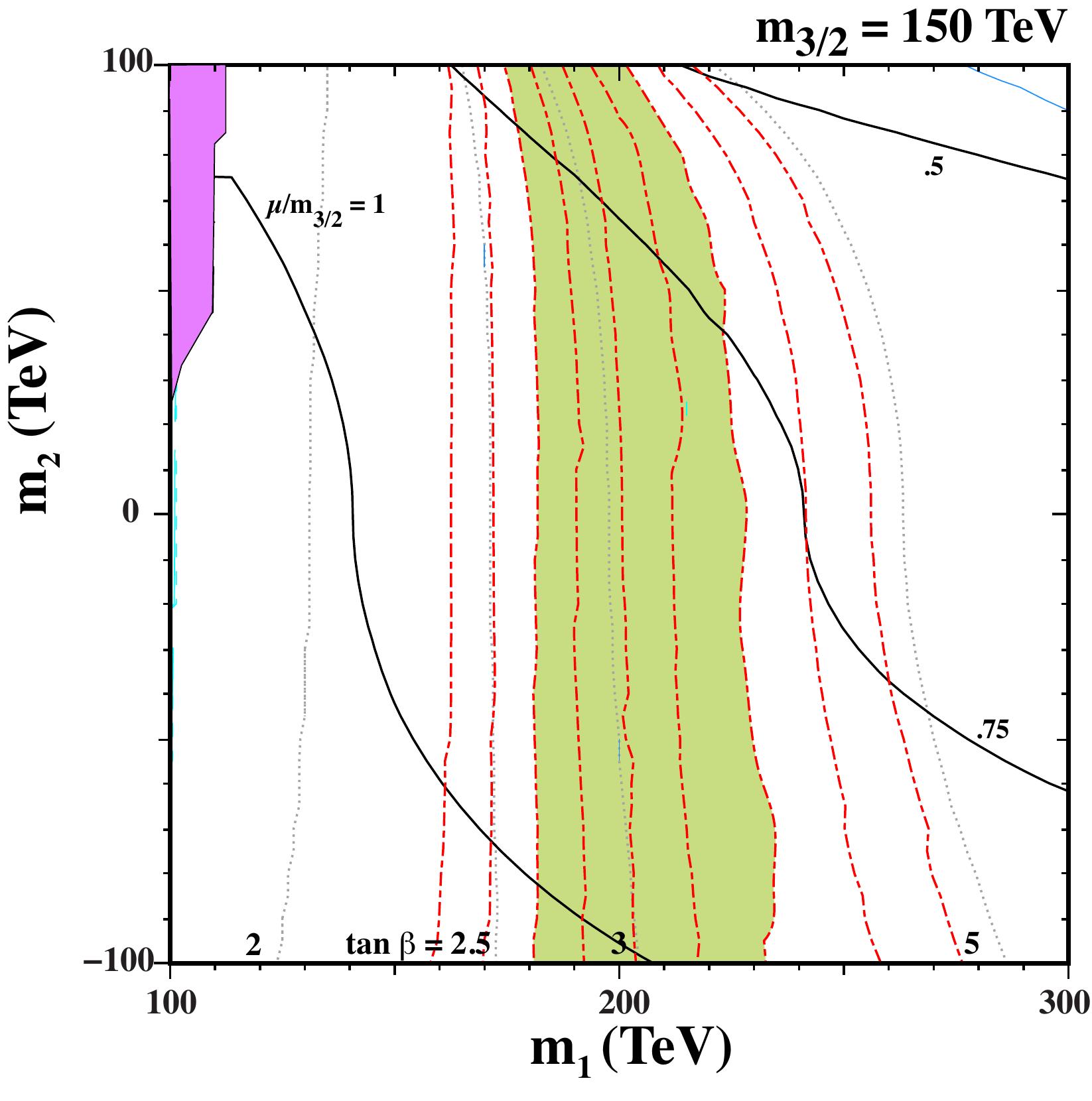,height=3.2in}
\hfill
\end{minipage}
\begin{minipage}{8in}
\epsfig{file=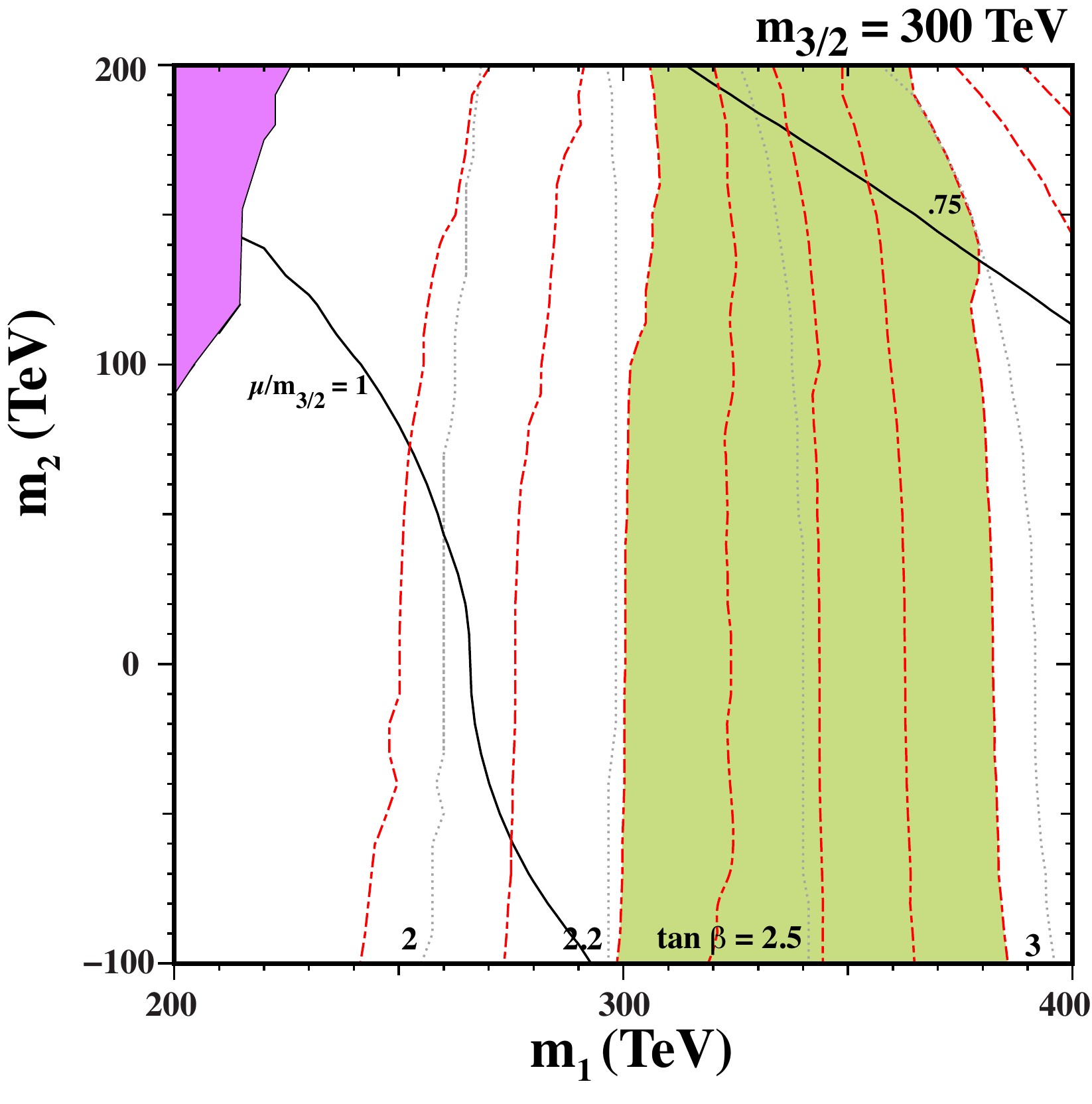,height=3.2in}
\epsfig{file=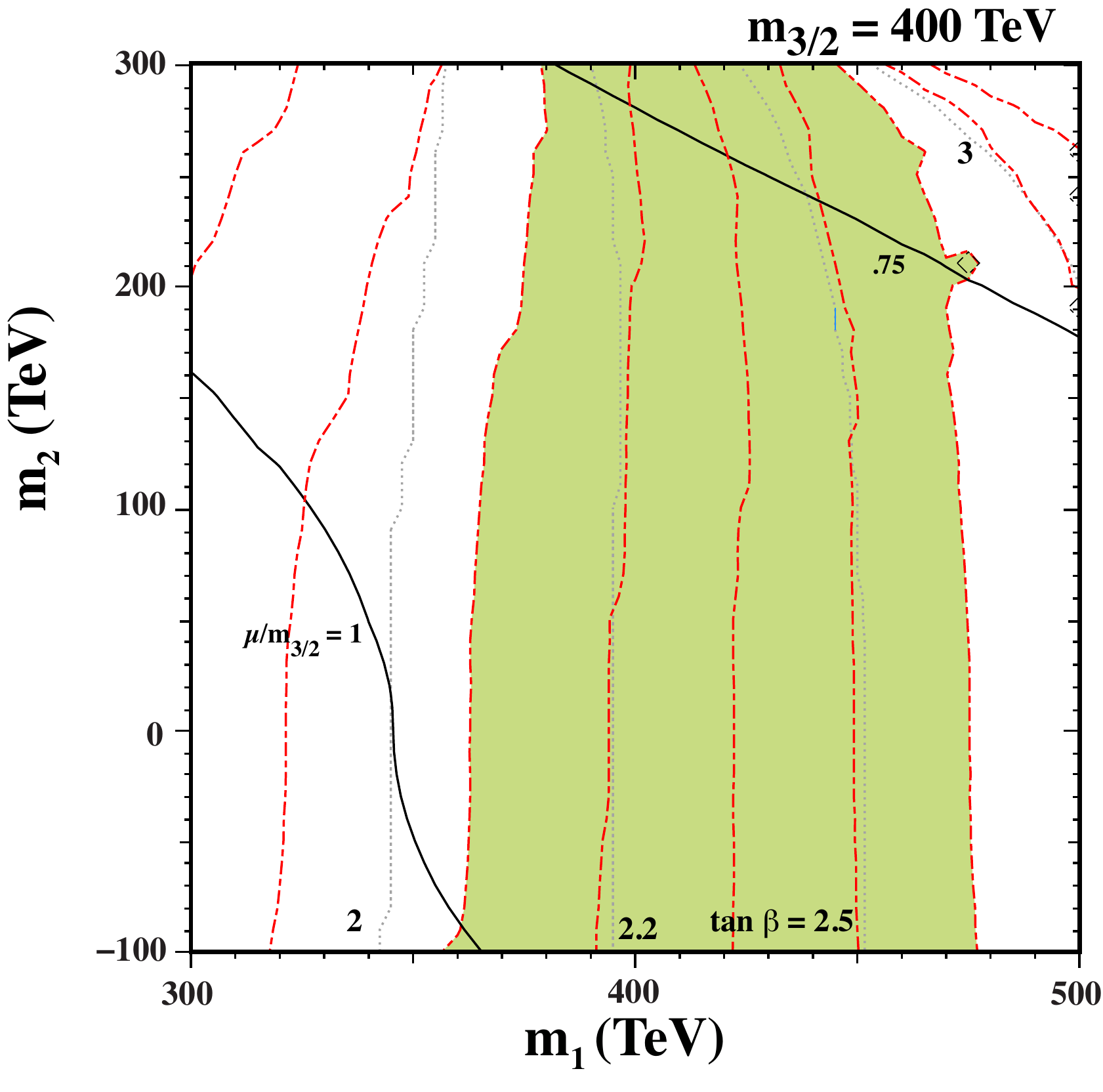,height=3.2in}
\hfill
\end{minipage}

\caption{
{\it
The $(m_1, m_2)$ plane for fixed $m_{3/2}$ = 60, 150, 300, and 400 TeV.
Shown are the contours for the light Higgs mass, $m_h$ (red, dot dashed) from 122 - 130 GeV
in 1 GeV intervals. The region with a Higgs mass between 124 and 128 GeV is shaded green.
The gray dotted curves show the calculated values of $\tan \beta$.  The range spans 2- 15 when
$m_{3/2} = 60$ TeV, and 2-3 when $m_{3/2} = 400$ TeV. Also shown are contours of $\mu/m_{3/2}$
(solid, black) which are typically close to 1. The blue shaded regions (when shown) correspond to
regions where no EWSB is possible.
}}
\label{m1m2}
\end{figure}

In Fig.~\ref{m1m2}, we show examples of the $m_1, m_2$ plane for fixed values of
$m_{3/2} = 60, 150, 300$, and 400 TeV. The red dot-dashed curves show contours of the
light Higgs mass, $m_h$ from 122 - 130 GeV in 1 GeV intervals.  The region
with 124 GeV $< m_h<$ 128 GeV is shaded green.  In all cases, we have
assumed the supergravity boundary condition of $B_0 = -m_{3/2}$ and $c_H = 0$, and
calculate $\mu$ and $\tan \beta$.  The solid black contours show
$\mu/m_{3/2}$ and as one can see, this ratio is close to one over much of the
displayed planes.  The exception occurs when $m_2^2$ is large and positive causing $\mu$ to become small.  In the figure, when $m_{3/2} = 60$ TeV, the region shaded blue at the top right of the figure
has $\mu^2 <0$ indicating the lack of an EWSB solution.  The region at low and negative $m_1^2$
is also excluded as there the Higgs pseudo-scalar mass, $m_A^2 < 0$.  Note the sign of the soft Higgs masses in the figure refers to the sign of the mass squared.  The gray dotted curves show the
calculated values of $\tan \beta$ which are typically around 4-5 when $m_{3/2} = 60$ TeV,
and are closer to 2 at larger $m_{3/2}$.

As one can see all of the viable solutions displayed in the figures require some degree
of non-universality.  In each case displayed, forcing Higgs mass universality would either require
a value of $m_1$ too small corresponding to a light Higgs with
mass $< 124$ GeV, or a value of $m_2$ too
large to allow EWSB solutions. As one can see, $m_1$ can be made universal, when
the gravitino mass is $\gtrsim 300$ TeV. For $m_{3/2} = 300$ TeV as shown in the lower
left panel, $m_1 = m_{3/2}$ is very close to the $m_h = 124$ GeV contour when $m_2$ is relatively small.  When $m_{3/2} = 400$ TeV as shown in the lower right panel, There is no
problem is obtaining a suitable Higgs mass. But for $m_2 \gtrsim 300$ TeV, $\mu^2$ quickly
runs negative and we lose the ability to satisfy the EWSB boundary conditions. At smaller, $m_2$
the results are in fact quite insensitive to the particular value of $m_2$. Therefore in what follows,
we will set $m_2 = 0$.

The vanishing of the up-type Higgs soft mass at the universality scale
can be explained \cite{eioy2} if either this Higgs field was part of a no-scale structure
\cite{noscale} of the  K\"ahler potential as in
\beq
K =   y y^* - 3 \log \left( 1 - \frac{1}{3}(H_2 H_2^*+ K^{(Z)}) \right) +  \log |W|^2
\label{K2}
\eeq
where $Z$ is the field(s) which breaks supersymmetry, and $y$ represents all other
fields including $H_1$. The resulting soft masses for the Higgs doublets in this case is
$m_2^2 = 0$. Alternatively, it is possible that the up-type doublet appears
as a Nambu-Goldstone boson described by the coset space, $U(3)/SU(2)\times U(1)$
\cite{Kugo:1983ai}.

\begin{figure}[h!]
\centering\epsfig{file=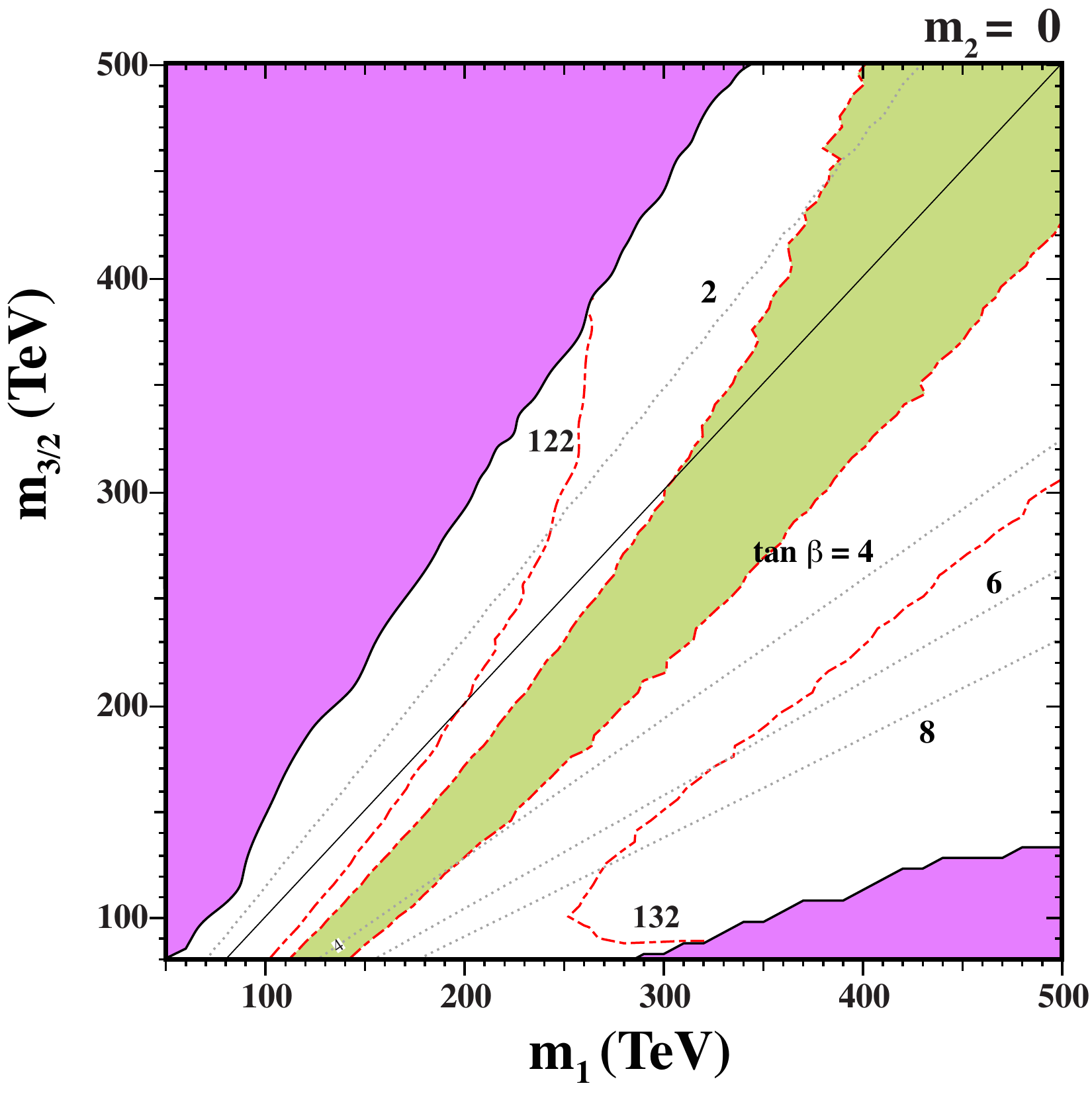,height=4in}

\caption{
{\it
The $(m_1, m_{3/2})$ plane for fixed $m_{2}$ = 0.
Shown are the contours for the light Higgs mass, $m_h$ (red, dot dashed) from 122 - 132 GeV
in 4 GeV intervals. The region with a Higgs mass between 124 and 128 GeV is shaded green.
The gray dotted curves show the calculated values of $\tan \beta$.  The black solid line shows
the down-type universal case, where $m_1 = m_{3/2}$.
 The blue shaded regions (when shown) correspond to
regions where no EWSB is possible.
}}
\label{m1m32}
\end{figure}

Having fixed $m_2 = 0$, it is possible to display the parameter space on
a single two-dimensional $m_1, m_{3/2}$ plane, as in Fig.~\ref{m1m32}.
As one can see, for low(er) values of $m_{3/2}$, the value of $m_1$ needed
to obtain $m_h$ between 124 and 128 GeV (shown as the shaded green region)
requires non-universality in the Higgs soft masses and $m_1 > m_{3/2}$.
When $m_{3/2} \gtrsim 300$ TeV, solutions with $m_1 = m_{3/2}$ become possible.
In all cases, we find $2 \lesssim \tan \beta \lesssim 4$.

\begin{figure}[h!]
\begin{minipage}{8in}
\epsfig{file=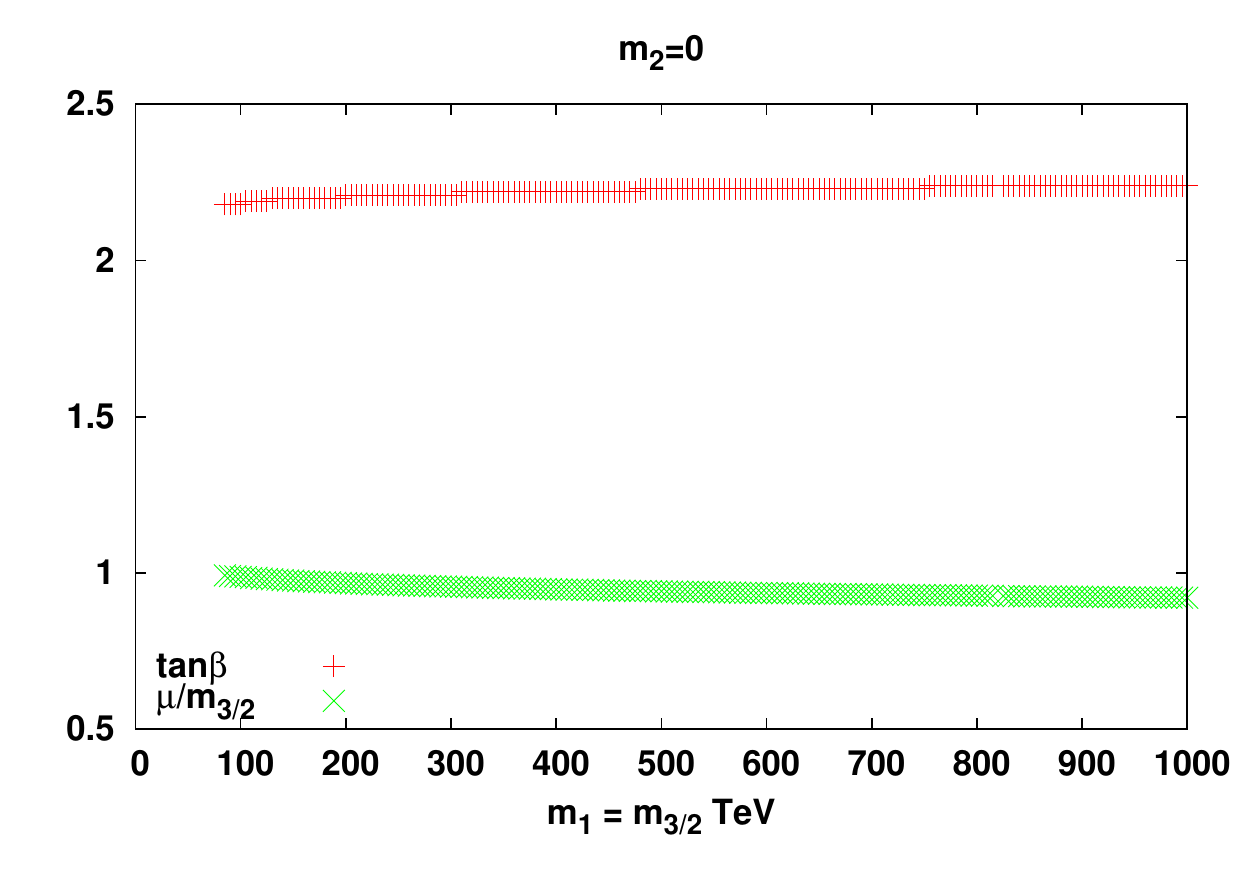,height=2.3in}
\epsfig{file=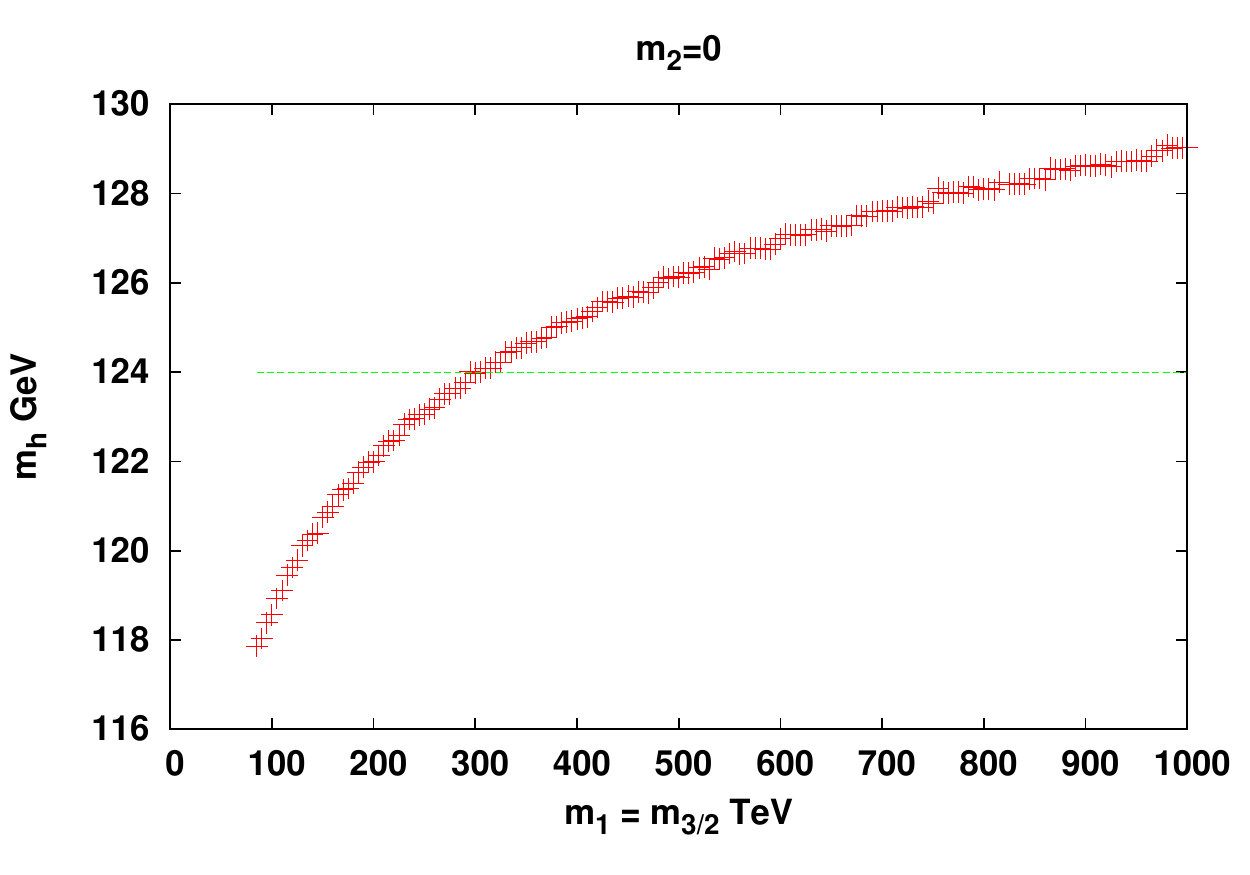,height=2.3in}
\hfill
\end{minipage}

\caption{Solutions for $\mu/m_{3/2}$ and $\tan \beta$ (left) and for $m_h$ (right)
in the one parameter version of the PQ-symmetric PGM model. Here $m_1 = m_{3/2}$
and $m_2 = 0$.
{\it
}}
\label{m1univ}
\end{figure}

We can go further and insist on universality of the down-type Higgs soft mass.
Results for this restrictive case are shown in Fig.~\ref{m1univ} where we
plot $\mu/m_{3/2}$, $\tan \beta$ (left) and $m_h$ (right) as a function of $m_1 = m_{3/2}$
for $m_2 = 0$.  As one can see, in this case, for a wide range of values of $m_{3/2}$,
$\mu/m_{3/2} \simeq 1$ and $\tan \beta \simeq 2.2$. However, in order to obtain
$m_h > 124$ GeV, we need 300 TeV $\lesssim m_{3/2} \lesssim 850 $ TeV.

\section{Universality With PQ-symmetry Breaking}
\label{sec:pquniv}
By generalizing the model for PQ-symmetry breaking, it may be possible to restore
full scalar mass universality.  The particular model we consider was first presented in \cite{Murayama:1992dj} and relates the PQ scale with see-saw scale for generating neutrino masses.  The superpotential for this model is
\begin{eqnarray}
W_{PQ}= \frac{f}{M_P}P^3Q+\frac{g}{M_P}H_u H_d PQ+\frac{1}{2}h_{ij}PN_iN_j \, ,
\end{eqnarray}
where $P,Q$ are the fields responsible for breaking the PQ-symmetry with charges $(-1,3)$, $N_i$ are the right handed neutrinos with PQ charge $1/2$, and $H_{u,d}$ have PQ charge $-1$.  In this model the right handed neutrino masses are generated by PQ-symmetry breaking, which is of order $10^{12}$ GeV as discussed above.  In pure gravity mediation models, these fields will also get supersymmetry breaking parameters.  The relevant supersymmetry breaking soft masses and $A$-terms  are
\begin{eqnarray}
-{\cal L}_{soft} \supset m_{Q}^2|Q|^2+m_P^2|P|^2+m_N^2|N|^2 + f\frac{m_A}{M_P}P^3Q +h.c.
\end{eqnarray}
with
\begin{eqnarray}
m_Q^2(\mu_{GUT})=m_P^2(\mu_{GUT})=m_N^2(\mu_{GUT})=m_{3/2}^2 \quad \quad m_A(\mu_{GUT})=m_{3/2}
\end{eqnarray}

The RG equations for these soft masses are dominated by the couplings $h_{ij}$.  In fact, if the number of neutrinos is large enough and the $h_{ij}$ are large enough, $m_P^2$ will be driven negative and the PQ-symmetry is broken. The vevs of $\langle P \rangle$ and $\langle Q \rangle $ will effectively be free parameters determined by $h_{ij}$ and $f$.   Breaking the PQ-symmetry in this manner will generate independent vevs for $\langle Q\rangle$ and $\langle P\rangle $. Because both $\langle P\rangle$ and $\langle Q \rangle$ are none zero, the $F$-terms for $P,Q$ will also be non-zero and independent.  With this set up, $\mu$ and $B\mu$ are linearly independent as in Eq. (\ref{eq:mu1}) and (\ref{eq:B1}) and in contrast to the case considered in Eq. (\ref{eq:mu2}) and (\ref{eq:B2}). This additional freedom in $\mu$ and $B\mu$ makes universal soft mass for the MSSM fields possible. For this scenario, the parameter space, defined by $m_{3/2}$ and $\tan\beta$, will be identical to that considered in \cite{eioy}.

There is another way to get universality for the scalar masses with the PQ breaking symmetry.  If we again include Giudice-Masiero mixing term for the $P$ and $Q$ in the K\"ahler potential, the relationship for $B$ of the Higgs bosons changes to
\begin{eqnarray}
B\mu=-m_{3/2}\left(1-2c_{PQ}\right)\mu \, .
\end{eqnarray}
Due to the freedom in $B$ from the $c_{PQ}$ term, we can independently define $\mu$ and $B$. This allows for solutions to the EWSB conditions to be found by solving for $k$ and $c_{PQ}$ leaving $\tan\beta$ as a free parameter\footnote{There will be some restrictions on this parameter space from the fact that $c_{PQ}>1$.}.

\section{Discussions}\label{sec:non-u}

The relatively large Higgs mass determined at the LHC coupled with the lack of discovery
of any superpartners indicates that the scale of supersymmetry must be higher than
originally thought if it is realized at low energy at all.  While it remains possible
that the discovery of supersymmetry is around the corner at scales
close to 1 TeV and well within reach on a 14 TeV LHC collider \cite{hope,ehow+},
it is also possible that the supersymmetry scale is significantly higher  and sits
in the range of 100 - 1000 TeV as expected in PGM models \cite{pgm,pgm2,pgm3,eioy,eioy2,eioy3},
models with strongly stabilized moduli \cite{klor,Dudas:2006gr,dlmmo}, or in so called models of
mini-split supersymmetry \cite{mini}.

In either case, we are forced to address the question regarding the scale
of supersymmetry breaking.  In models where the SUSY scale is upwards of 100 TeV,
gaugino masses are generally generated at the 1-loop level through anomalies. In that case,
the lightest supersymmetric particle is usually the wino.  For $\mu > 0$, the wino relic density
is too small to account for the dark matter of the universe, but the wino might be
a viable candidate when $\mu < 0$ and the SUSY scale is of order 500 TeV (see e.g. \cite{eioy}).
However, even in that case, there are strong constraints against wino dark matter from
higher energy gamma-ray observations \cite{gamma}. In this context the axion becomes
an attractive dark matter candidate.

In addition to the dark matter problems, most low energy supersymmetric models
suffer from the $\mu$-problem.  Even in models where the scale of supersymmetry
breaking is generated spontaneously, the $\mu$ term, being supersymmetric, is typically
put in by hand as a bilinear in the superpotential. Therefore, the dynamical generation of the
$\mu$ term, through the coupling of the MSSM Higgs doublets to Standard Model
singlets with non-zero PQ charge \cite{KN} offers an attractive solution to potentially both
the $\mu$ term and dark matter problems.

From Eq. (\ref{relic}), it is clear that a PQ scale of close to $10^{12}$ GeV is needed, if
axions are to be the dominant form of dark matter in the universe.
Thus if axions make up any significant component of the dark matter  (say,
at least 10\% of the dark matter), the $\mu$ term is expected to be at least several TeV.
If axions are the dominant form of dark matter, then the $\mu$ term is expected to be
of order 100 TeV or more.

In this paper, we have shown how a model of PGM can be constructed
in the context of supergravity with a PQ coupling to the MSSM
Higgs doublets. Because of this coupling, a Giudice-Masiero like term in the
K\"ahler potential is not allowed and we must deviate slightly
from pure scalar mass universality. Allowing for the possibility of non-universal
Higgs masses, we have shown that viable models exist with (non-Higgs)
scalar mass universality which respect minimal supergravity boundary
conditions for the $B$-term.  Gaugino masses and $A$-terms are
assumed to arise from anomaly mediated contributions.
All soft terms and couplings are run down from the universality scale (assumed to be
the GUT scale) and minimization of the Higgs potential is used to determine the $\mu$ term and
$\tan \beta$. If axions are the dominant form of dark matter, the coupling of the
Higgs doublets to the PQ fields thus generates a $\mu$ term of order of 100 TeV
which sets the scale for supersymmetry breaking.
While it is possible to find solutions with $\mu \ll m_{3/2}$ (for example when $m_2$ is large
and $\mu^2$ is driven to 0), it is not possible to find solutions with $\mu \gg m_{3/2}$.
Thus fixing $\mu \gtrsim \mathcal{O}(100)$ TeV in order to obtain a significant axion relic density
forces us into the domain where supersymmetry is broken at a similarly high scale\footnote{ See \cite{why} for a similarly motivated model.}.

Fortuitously, it also possible to find solutions of the type just described
with a Higgs mass in the range determined at the LHC.  For relatively low
$m_{3/2} \lesssim 300$ TeV, we require $m_1 \gtrsim m_{3/2}$. While for larger
$m_{3/2}$ up to 850 TeV, solutions with $m_1 = m_{3/2}$ are possible.  Our results are not
particularly sensitive to $m_2$, though $m_2 < m_{3/2}$ is quite generic.
If the up-type Higgs is a pseudo Nambu-Goldstone boson or is part of a
no-scale structure so that $m_2 = 0$, we are left with a particularly simple model
with one free parameter, $m_{3/2}$. For $m_{3/2} > 300$ TeV, we have
$m_h > 124$ GeV, and $\mu/m_{3/2} \simeq 1$ and $\tan \beta \simeq 2.2$.

We have also shown several mechanisms which suppress isocurvature fluctuations
despite having a dominant component of axion dark matter.
The simplest possibility, which may not apply here, is that described in \cite{al91} and requires only
a small coupling $\lambda$ in Eq. (\ref{wpq2}).  The constraint on $\lambda$ can be significantly relaxed if a Giudice-Masiero-like term is added to the K\"ahler potential as in Eq.~(\ref{cpq}).
Finally, we have proposed a novel dynamical mechanism for the generation of the
axion decay constant which allows $F_{PQ} \simeq M_P$ during inflation,
and smaller values ($\mathcal{O}(10^{12})$ GeV) at low energy.

\section*{Acknowledgments}
The work of J.E. and K.A.O. was supported in part
by DOE grant DE--FG02--94ER--40823 at the University of Minnesota.
This work is also supported by Grant-in-Aid for Scientific research from the
Ministry of Education, Science, Sports, and Culture (MEXT), Japan, No.\ 22244021 (T.T.Y.),
No.\ 24740151 (M.I.), and also by the World Premier International Research Center Initiative (WPI Initiative), MEXT, Japan.

\end{document}